\documentclass[twocolumn,aps,prl,superscriptaddress]{revtex4}
\usepackage{graphicx}
\usepackage{bm}

\def\bea{\begin{eqnarray}}
\def\eea{\end{eqnarray}}
\def\be{\begin{equation}}
\def\ee{\end{equation}}

\begin{document}

\title{Narrow Line Photoassociation in an Optical Lattice}

\author{T. Zelevinsky}
\affiliation{JILA, National Institute of Standards and Technology and
University of Colorado, \\
and the Department of Physics, University of Colorado, Boulder, CO 80309-0440, USA}
\author{M. M. Boyd}
\affiliation{JILA, National Institute of Standards and Technology and
University of Colorado, \\
and the Department of Physics, University of Colorado, Boulder, CO 80309-0440, USA}
\author{A. D. Ludlow}
\affiliation{JILA, National Institute of Standards and Technology and
University of Colorado, \\
and the Department of Physics, University of Colorado, Boulder, CO 80309-0440, USA}
\author{T. Ido}
\affiliation{JILA, National Institute of Standards and Technology and
University of Colorado, \\
and the Department of Physics, University of Colorado, Boulder, CO 80309-0440, USA}
\affiliation{PRESTO, Japan Science and Technology Agency, 4-1-8 Honcho,
Kawaguchi, 332-0012, Japan}
\author{J. Ye}
\affiliation{JILA, National Institute of Standards and Technology and
University of Colorado, \\
and the Department of Physics, University of Colorado, Boulder, CO 80309-0440, USA}
\author{R. Ciury{\l}o}
\affiliation{Instytut Fizyki, Uniwersytet Miko\l aja Kopernika, ul. Grudzi\c{a}dzka
5/7, 87-100 Toru\'{n}, Poland}
\author{P. Naidon}
\author{P. S. Julienne}
\affiliation{Atomic Physics Division, National Institute of Standards and Technology,
Gaithersburg, MD 20899-8423, USA}


\begin{abstract}     
With ultracold $^{88}$Sr in a 1D magic wavelength optical lattice,
we performed narrow line photoassociation spectroscopy near the
$^1$S$_0 - ^3$P$_1$ intercombination transition.  Nine least-bound
vibrational molecular levels associated with the long-range $0_u$
and $1_u$ potential energy surfaces were measured and identified. A
simple theoretical model accurately describes the level positions
and treats the effects of the lattice confinement on the line
shapes. The measured resonance strengths show that optical tuning of
the ground state scattering length should be possible without
significant atom loss.

PACS numbers: 34.80.Qb, 32.80-t, 32.80.Cy, 32.80.Pj

\end{abstract}
\date{\today}
\maketitle

\newcommand{\w}{3.25in}

\newcommand{\PASchematicFigure}[1][\w]{
\begin{figure}
\includegraphics*[width=2.6in]{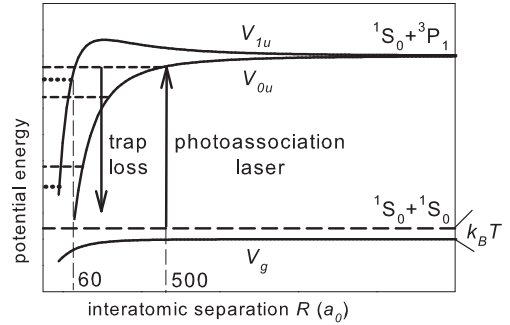}\hfill
\caption{Schematic diagram of the long-range Sr$_2$ molecular
potentials (not to scale).  The ground and excited molecular states
coincide with two ground state atoms, and with one ground and one
excited state atom, respectively, at large internuclear separations.
The ground state has $gerade$ symmetry and its energy is given by
the potential $V_g$, while the excited state $ungerade$ potentials
that support dipole transitions to the ground state are $V_{0u}$ and
$V_{1u}$, the latter with a small repulsive barrier.  All
vibrational states of $0_u$ and $1_u$ (dashed and
dotted lines, respectively) are separated by more than the natural
line width, permitting high resolution PA
spectroscopy very close to the dissociation limit when the atoms are
sufficiently cold.} \label{fig:PASchematic}
\end{figure}
}

\newcommand{\ExptSetupFigure}[1][\w]{
\begin{figure}
\includegraphics*[width=3.0in]{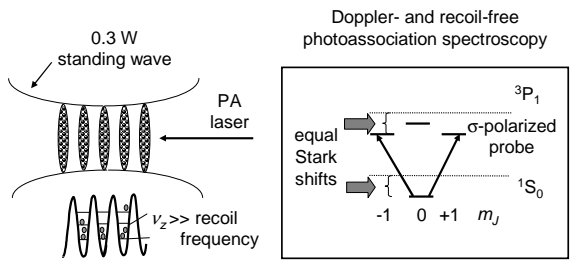}\hfill
\caption{Experimental configuration.  The 2 $\mu$K atoms are confined in a 1D optical
lattice in the Lamb-Dicke regime that allows Doppler- and recoil-free photoassociation
spectroscopy.  The lattice laser intensity is large enough to
confine most atoms in
the ground motional state.  The 689 nm weak PA laser co-propagates with
the lattice.  The 914 nm lattice wavelength ensures that
$^1$S$_0$ and the $m_J = \pm 1$ sublevels of $^3$P$_1$
are Stark-shifted by equal amounts.}
\label{fig:ExptSetup}
\end{figure}
}

\newcommand{\SpectrumFigure}[1][\w]{
\begin{figure}
\includegraphics*[width=3.0in]{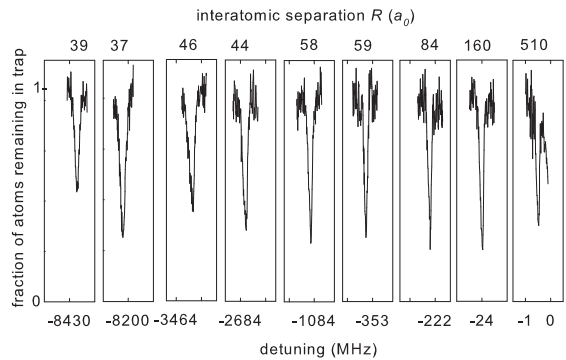}\hfill
\caption{The spectra of the long-range Sr$_2$ molecule near the
$^1$S$_0 - ^3$P$_1$ dissociation limit.  The horizontal scale is marked
on the rightmost panel; different PA laser intensities were used for
each spectrum due to largely varying line strengths.  The top labels
indicate the interatomic separations that correspond to the classical
outer turning points of each resonance.}
\label{fig:PASpectrum}
\end{figure}
}

\newcommand{\TempFigure}[1][\w]{
\begin{figure}
\includegraphics*[width=3.0in]{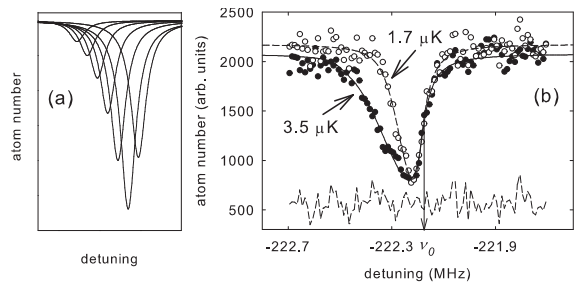}\hfill
\caption{(a) A line is schematically shown as
a sum of Lorentzians with positions and amplitudes given by
a continuous thermal distribution.
(b) The measured line shapes of the $-222$ MHz transition at 1.7
$\mu$K (open circles) and 3.5 $\mu$K (filled circles), clearly showing the
effect of temperature on the red side of the line.  The bottom trace
shows the residuals of the 1.7
$\mu$K curve fit.}
\label{fig:Temp}
\end{figure}
}

\newcommand{\IntensityFigure}[1][\w]{
\begin{figure}
\includegraphics*[width=2.5in]{ProbeInt1084MHz.eps}\hfill
\caption{The $-1084$ MHz line spectrum taken at three different PA laser
intensities.  The depths of the photoassociation lines are
key to determining the optical length of each
resonance.}
\label{fig:ProbeInt}
\end{figure}
}

Photoassociation (PA) spectroscopy
\cite{BohnPA99,JulienneReviewInPress} is a valuable tool for studies
of atomic collisions and long-range molecules
\cite{PhillipsNaPA02,TannoudjiHePA03,
TiemannCaPA03,TakahashiYbPA04,KillianPA05,KatoriScatLengths,KillianScatLengths05}.
In contrast to conventional molecular spectroscopy that probes the
most deeply bound vibrational levels of different electronic states,
PA spectroscopy excites the vibrational levels close to the
dissociation limit, where the molecular properties depend mostly on
the long-range dipole-dipole interatomic interactions. Molecular
potentials are much simpler at long range than at short range, and
are directly related to the basic atomic properties such as the
excited state lifetime and the $s$-wave scattering length.

In this Letter, we report the use of PA spectroscopy to resolve
nine least-bound vibrational levels of the $^{88}$Sr$_2$ dimer near
the $^1$S$_0 - ^3$P$_1$ intercombination line.  Our identification
of these levels is required for tuning of the
ground state scattering length with low-loss optical Feshbach
resonances \cite{JulienneOptFesh05}.  This is of
great interest for Sr, since magnetic Feshbach resonances are absent
for the ground state, and the background scattering length is too
small to allow evaporative cooling \cite{KillianScatLengths05}.
In contrast to prior PA work that utilizes strongly allowed transitions
with typical line widths in the MHz range, here the spin-forbidden
$^1$S$_0 - ^3$P$_1$ line has a natural width of 7.5 kHz.  This
narrow width allows us to measure the least-bound vibrational levels
that would otherwise be obscured by a broad atomic line, and
to observe characteristic thermal line shapes even at $\mu$K atom
temperatures. It also permits examination of the unique crossover
regime between the van der Waals and dipole-dipole interactions,
that occurs in Sr near the $^1$S$_0 - ^3$P$_1$
dissociation limit.  This access to the
van der Waals interactions ensures large bound-bound Franck-Condon
factors, and may lead to more efficient creation of cold ground state Sr$_2$
molecules with two-color PA than what is possible using broad
transitions.  Bosonic $^{88}$Sr is
ideal for narrow line PA because of its
hyperfine-free structure and a high isotopic abundance (83\%).
In addition, Sr is
a candidate atom for optical clocks
\cite{Ido88SpectrPRL05,TakamotoNature05,LudlowPRL06},
and PA spectroscopy
could help determine density shifts for various clock transitions.

To enable long interrogation times
needed for narrow line photoassociation, we trap Sr atoms in a 1D
optical lattice.  This technique also
suppresses recoil shifts and Doppler broadening
and thus simplifies the PA line shapes. The specially chosen
lattice wavelength
$(\lambda_{\rm{magic}})$ induces state-insensitive AC Stark shifts that
minimize the perturbing effects of the lattice
\cite{IdoFirstLatticePRL03}.

\PASchematicFigure Figure \ref{fig:PASchematic} illustrates the
relevant potential energy curves for the Sr$_2$ dimer as a function
of interatomic separation $R$.  The photoassociation laser at 689 nm
induces allowed transitions between the separated $^1$S$_0$ atom
continuum at a temperature $T$ to the vibrational bound levels of
the excited $ungerade$ potentials $V_{0u}$ and $V_{1u}$,
corresponding to the total atomic angular momentum projections onto
the internuclear axis of 0 and 1, respectively.  Although
our calculations, based on
the coupled channels method in Ref.~\cite{JulienneNarrowLinePA04},
account for the Coriolis mixing  between the $0_u$ and
$1_u$ states, this mixing is small enough that $0_u$ and $1_u$ are good
symmetry labels for classifying the states.
Including the Coriolis mixing is necessary in order to fit the
observed $0_u$ and $1_u$ bound state energies.  Only $s$-waves
(total ground state molecular angular momentum
$J=0$) contribute to the absorption (excited state $J=1$) at our low
collision energies.  The long-range potentials are given by the $C_6/R^6$ van
der Waals and $C_3/R^3$ dipole-dipole interactions, plus a
rotational term
$\propto 1/R^2$,
\bea \label{eq:V0u}
V_{0u} &= -C_{6,0u}/R^6-2C_3/R^3+h^2 A_{0u}/(8\pi^2\mu R^2), \\
V_{1u} &= -C_{6,1u}/R^6+C_3/R^3+h^2 A_{1u}/(8\pi^2\mu R^2),
\label{eq:V1u} \eea
where $h$ is the Planck constant, $\mu$ is the reduced mass,
$A_{0u}=J(J+1) +2$, and $A_{1u}=J(J+1)$.  The values of $C_3$,
$C_{6,0u}$,
and $C_{6,1u}$ are adjusted in our theoretical model so that
bound states exist at the experimentally determined resonance energies.
Since our modeling of the PA energy levels due to long-range
interactions is
not sensitive to the details of short-range interactions, the latter are
represented by simple 6-12 Lennard-Jones potentials whose depths are
chosen to approximately match those in Ref. \cite{ShortRangeSrCPL03}.
The coefficient $C_3$ can be expressed in terms of the atomic
lifetime $\tau$ as $C_3 = 3\hbar c^3/(4\tau\omega^3)$, where
$\hbar\omega$ is the atomic transition energy and $c$ is the speed
of light. Since the $C_3$ and $C_6$ interactions have comparable
magnitudes in the region of $R$ relevant to the energy levels we
observe,
the standard semiclassical analysis based on a single interaction
$\propto R^{-n}$ \cite{LeRoy70} is not applicable.

\ExptSetupFigure The experiment is performed with $\sim 2$ $\mu$K
$^{88}$Sr atoms that are
trapped and cooled in a two-stage magneto-optical trap (MOT)
\cite{LoftusPRL04,LoftusPRA04}.
As the atoms are cooled in the 689 nm narrow line
$^1$S$_0 - ^3$P$_1$ MOT, they are
loaded into a 1D far-detuned optical lattice
\cite{LudlowPRL06}, which is a 300 mW standing wave of 914 nm light
with a 70 $\mu$m beam waist at the MOT.  This results in $10^5$
atoms with the average density of about $3\times 10^{12}$/cm$^3$,
after the MOT is switched off.  The axial
trapping frequency in the lattice is $\nu_z\sim 50$ kHz, much larger than the 5
kHz atom recoil frequency and the 7.5 kHz line width,
resulting in Doppler- and recoil-free
spectroscopy if the PA laser is collinear with the lattice
beam.  Figure \ref{fig:ExptSetup} illustrates the lattice
configuration.  The polarizations of the lattice and PA lasers are
linear and orthogonal in order to satisfy the
$\lambda_{\rm{magic}}$ condition at 914 nm
\cite{IdoFirstLatticePRL03}. The
choice of $\lambda_{\rm{magic}}$ ensures that $^1$S$_0$ and the magnetic sublevels
of $^3$P$_1$ that are coupled to $^1$S$_0$ by the PA laser are
Stark shifted by equal amounts, resulting in the zero net Stark
shift for the transition and consequently eliminating inhomogeneous
Stark broadening and shifts of the PA lines.  A 689 nm diode laser
used for PA is offset-locked to a cavity-stabilized
master laser with
a sub-kHz spectral width \cite{Ido88SpectrPRL05,LudlowPRL06}.

\SpectrumFigure
To trace out the molecular spectra, the PA laser
frequency is stepped, and after 320 ms of photoassociation at a fixed
frequency the atoms are released from the lattice and
illuminated with a 461 nm light pulse (resonant with $^1$S$_0 - ^1$P$_1$)
for atom counting.
At a PA resonance, the atom number drops as excited
molecules form and subsequently decay to ground state molecules in high vibrational
states or hot atoms that cannot remain trapped.
Figure \ref{fig:PASpectrum} shows the nine observed PA line spectra
near the dissociation limit.

The line fits for the free-bound transitions are based on a convolution
of a Lorentzian profile with a thermal distribution of initial
kinetic energies \cite{JonesLineShape99}, as
schematically shown in Fig. \ref{fig:Temp} (a).  In the 1D optical lattice,
we work in the quasi-2D scattering regime, since
$T\sim 2$ $\mu$K $\sim \nu_z/k_B$ ($k_B$ is the Boltzmann constant),
and the axial motion is quantized.  For weak PA
laser intensities used here, the total
trap loss rate is a superposition of loss rates $K_{\varepsilon}$ for atoms
with thermal energies $\varepsilon\equiv h^2k^2/(8\pi^2\mu)$,
\be
K_{\varepsilon}(\nu) = \frac{h}{\mu}\frac{\gamma^2 l_{\rm{opt}}}{(\varepsilon/h+\nu-\nu_0)^2+(\gamma+\gamma_1)^2/4},
\label{eq:Kepsilon}
\ee
where $\nu$ and $\nu_0$ are the laser and molecular resonance frequencies
expressed as detunings from the atomic $^1$S$_0 - ^3$P$_1$ line, $\gamma\simeq 15$
kHz is the line width of the excited molecular state, $\gamma_1\simeq 25$ kHz
is a phenomenological parameter that accounts for the observed
line broadening, $l_{\rm{opt}}$ is the
optical length \cite{JulienneOptFesh05} discussed below, and only $s$-wave
collisions are assumed to take place.  While the
total 3D PA loss rate is
\be
K_{3D}(T, \nu) = \frac{2}{\sqrt{\pi}}\int_{0}^{\infty}{K_{\varepsilon}(\nu)e^{-\varepsilon/k_BT}\frac{\sqrt{\varepsilon}d\varepsilon}{(k_BT)^{3/2}}},
\label{eq:KT_3D}
\ee
the 2D loss rate is
\be
K_{2D}(T, \nu) = \int_{0}^{\infty}{K_{\varepsilon}(\nu+\nu_z/2)e^{-\varepsilon/k_BT}\frac{d\varepsilon}{k_BT}}.
\label{eq:KT_2D}
\ee
Dimensional effects appear in Eq. (\ref{eq:KT_2D}) as a larger density of states
at small thermal energies ($i.e.$ no $\sqrt{\varepsilon}$ factor), and as a red shift
by the zero-point confinement frequency $\nu_z/2$.  Another property of a 2D
system (not observed here due to small probe intensities) is a lower bound on power
broadening near $T\sim 0$, fixed by the zero-point momentum \cite{Petrov2D01}.
Equation (\ref{eq:KT_2D}) assumes only collisions in the motional
ground state of the lattice potential.  Although 50\% of our atoms
are in excited motional states, using a more complete expression yielded
essentially the same line shapes.  If any local atom
density variations associated with PA are neglected and the temperature is
assumed to be constant during probing, then the atom density evolution is given by
\be
\dot{n} = -2Kn^2-n/\tau_l,
\label{eq:DensityDiffEq}
\ee
where $\tau_l \sim 1$ s is the lattice lifetime, from which we calculate the
number of atoms remaining in the lattice.

The measured PA resonances $\nu_0$ were reproduced theoretically
by adjusting the $C_3$ and $C_6$ parameters as well as
the short-range interatomic potentials.
The measured and calculated $\nu_0$
are compared in Table \ref{table:Table1}.
\begin{table}[h]
\begin{tabular}{|r|r||r|r|} \hline
$0_u$ Measured & $0_u$ Calculated & $1_u$ Measured & $1_u$ Calculated \\
(MHz) & (MHz) & (MHz) & (MHz) \\ \hline
-0.435(37) & -0.418 & -353.236(35) & -353.232 \\
-23.932(33) & -23.927 & -2683.722(32) & -2683.729 \\
-222.161(35) & -222.152 & -8200.163(39) & -8112.987 \\
-1084.093(33) & -1084.091 & & \\
-3463.280(33) & -3463.296 & & \\
-8429.650(42) & -8420.124 & & \\ \hline
\end{tabular}
\caption{The comparison of measured and calculated molecular resonance
frequencies, for both $0_u$ and $1_u$ potentials.}
\label{table:Table1}
\end{table}
The positions of the measured lines are
found by fitting Eqs. (\ref{eq:Kepsilon}-\ref{eq:DensityDiffEq})
to the data (see Fig. \ref{fig:Temp}).
Both 2D and 3D line shape models were applied, without an appreciable
difference in the obtained optical lengths.  However, the 2D
model yields $\nu_0$ corrections of $+15(10)$ kHz
that are included in Table \ref{table:Table1}.
The measurement errors in the Table are a combination of a 30 kHz uncertainty
of the atomic reference for the PA laser, and of
the systematic and statistical reproducibility.
The calculated $\nu_0$ results are based on the coupled channels
model \cite{JulienneNarrowLinePA04} and were obtained using
$C_3 = 0.007576$ au ($\tau = 21.35$ $\mu$s), $C_{6,0u} = 3550$ au, and
$C_{6,1u} = 3814$ au (1 au is $E_h a_0^3$ for $C_3$ and $E_h a_0^6$ for
$C_6$, where $E_h = 4.36\times 10^{-18}$ J and $a_0 = 0.0529$ nm).
However, the model limitations, such as insufficient
knowledge of the short-range potentials, require the following
uncertainties in the long-range parameters:
$C_3 = 0.0076(1)$ au ($\tau = 21.4(2)$ $\mu$s), $C_{6,0u} = 3600(200)$ au, and
$C_{6,1u} = 3800(200)$ au.
The theoretical results for the two PA lines at detunings
over 8 GHz disagree with experiment by 0.1-1\% due to a high sensitivity
to the short-range potentials, while all the other lines fit well
within the specified uncertainties of as low as $10^{-5}$.

\TempFigure
Since the width of the Sr intercombination line is in the kHz range, even
ultracold temperatures of a few $\mu$K contribute significant thermal
broadening.  Figure \ref{fig:Temp} (b) shows two spectra of the $-222$ MHz
PA line taken at $T=1.7$ $\mu$K and $T=3.5$ $\mu$K ($T$ was measured
from the time-of-flight atom cloud expansion).  The spectrum
of the hotter sample manifests thermal broadening as a tail
on the red-detuned side.

A photoassociation line depth is given by the
optical length \cite{JulienneOptFesh05}, $l_{\rm{opt}}$, a parameter
proportional to the free-bound Franck-Condon factor and the PA laser
intensity.  It is useful for specifying the laser-induced changes in
both elastic and inelastic collision rates that
lead to optical control of the atomic scattering length
and to atom loss, respectively.  The optical lengths were obtained from the data
and calculated from the Franck-Condon factors under the assumption
of a zero ground state background
scattering length \cite{KillianScatLengths05}.  For the
$-0.4$ MHz line, the experimental $l_{\rm{opt}}/I\simeq 4.5\times 10^5$
$a_0/$(W/cm$^2$), while
the theoretical value is $9.0\times 10^5$ $a_0/$(W/cm$^2$), where $I$ is the
PA laser intensity.  Significant
experimental uncertainties may arise from
the calibration of the atomic density and the focused PA laser intensity.
In addition, we observe line broadening $\gamma_1$, most likely a
result of the residual magnetic field, and any variation in this extra
width can lead to $l_{\rm{opt}}$ error.  The $-0.4$ MHz resonance is
special due to its very large classical turning point, and its
effective optical length increases with decreasing $\varepsilon$,
while $l_{\rm{opt}}$ is energy-independent for
the other lines (due to the Wigner threshold law \cite{Wigner48}).
Therefore, the reported effective
$l_{\rm{opt}}$ values for the $-0.4$ MHz line are based on the 2 $\mu$K
sample temperature with the 1 $\mu$K zero-point energy.  In addition,
the fitted $l_{\rm{opt}}$ for this resonance was multiplied by 2.5 to obtain the
value quoted above,
since our calculations show that 60\% of the photoassociated and decayed atoms
have insufficient kinetic energies to leave the $\sim 10$ $\mu$K deep trap.
The measured and calculated $l_{\rm{opt}}$ for the other PA lines can
differ by up to an order of magnitude, but both decrease rapidly as the
magnitude of the detuning increases and are of the order of
$l_{\rm{opt}}/I=1$ $a_0/$(W/cm$^2$) for the most deeply bound levels.
The theoretical optical length of the $-222$ MHz line was found to have
the largest dependence on the assumed ground state scattering length due to
its sensitivity to the nodal structure of the ground state wave function.

Intercombination transitions of alkaline earths
such as Sr are particularly good candidates for optical control
of the ground state scattering length, $a_{\rm{opt}}$,
because there is a possibility of large gains in $a_{\rm{opt}}$ with
small atom losses.
In fact, using the $-0.4$ MHz PA line with the measured
$l_{\rm{opt}}\sim 5\times 10^5$ $a_0$ cm$^2$/W
will allow tuning the ground state
scattering length by
\cite{JulienneOptFesh05} $a_{\rm{opt}}=l_{\rm{opt}}(\gamma/\delta)f\simeq \pm 300$
$a_0$, where the PA laser with $I=10$ W/cm$^2$ is far-detuned by
$\delta = \pm 160$ MHz from the molecular resonance,
and the factor
$f=(1+(\gamma/2\delta)^2(1+2kl_{\rm{opt}})^2)^{-1}\simeq 0.8$
accounts for the power broadening for 3 $\mu$K collisions.
In contrast, optical tuning of the scattering length in alkali
$^{87}$Rb \cite{GrimmOptFeshRb04} achieved $a_{\rm{opt}} = \pm 90$ $a_0$
at much larger PA laser intensities of 500 W/cm$^2$.  In addition, the Sr
system  at the given parameter values will have a loss rate  of
\cite{JulienneOptFesh05}
$K=(2h/\mu)a_{\rm{opt}}\gamma/(2\delta)\simeq 2\times10^{-14}$
cm$^3$/s, while the loss rate in the $^{87}$Rb experiment was
$2\times10^{-10}$ cm$^3$/s.
The overall optical length gain of over 5 orders of magnitude
is possible for the Sr system because the narrow intercombination
transition allows access to the least-bound molecular state, and
the PA line strength increases with decreasing detuning from
the atomic resonance (see Fig. 1 in Ref. \cite{JulienneOptFesh05}).
The above $a_{\rm{opt}}$ and $K$ values accessible with the
$-0.4$ MHz resonance result in the elastic and
inelastic collision rates of
$\Gamma_{\rm{el}}\sim\sqrt{(2\varepsilon/\mu)}8\pi a_{\rm{opt}}^2 n\sim 600$/s
and $\Gamma_{\rm{inel}}\sim 2Kn\sim 0.1$/s, respectively.  The favorable
$\Gamma_{\rm{el}}/\Gamma_{\rm{inel}}$ ratio may enable
evaporative cooling.
A stringent constraint on the use of narrow-line optical Feshbach resonances is the
proximity of the atomic transition.  In the above example, the
scattering rate of the PA laser photons due to the $^1$S$_0 - ^3$P$_1$
atomic line is $\Gamma_s\sim 40$/s $\sim\Gamma_{\rm{el}}/15$.
Excessive one-atom photon scattering can cause heating and loss
of atoms from the lattice.

Another application of narrow line PA is potentially efficient production of
cold molecules in the ground state.  Our bound-bound Franck-Condon factor
calculations show that over 50\% of the molecules electronically
excited to the $-222$ MHz level decay to the last ground state vibrational
level (distributed between $J = 0,2$).

In summary, we have measured a series of nine vibrational levels
of the $^{88}$Sr$_2$ dimer near the $^1$S$_0 - ^3$P$_1$ intercombination
transition.  The 20 $\mu$s long lifetime of $^3$P$_1$, combined with the magic
wavelength optical lattice technique,
allowed direct probing of the least-bound state, and observation of
thermal photoassociation line shapes even at ultracold $\mu$K temperatures,
with explicit dimensional effects.
We have characterized the strengths of the molecular resonances, and shown
that the least-bound state should allow a unique combination of extensive, yet
low-loss optical control of the $^{88}$Sr ground state scattering length
that may enable efficient evaporative cooling.

We thank N. Andersen for stimulating discussions, and acknowledge financial
support from NIST, NRC, NSF, ONR, and the National Laboratory FAMO in Toru\'{n}.



\end{document}